\documentstyle[12pt]{article}
\newcommand{\bm}[1]{\mbox{\boldmath$#1$}}
\newcommand{\lw}[1]{\smash{\lower2.0ex\hbox{#1}}}
\begin{document}
\vspace{2cm}
\title{Magnetic susceptibility and low-temperature specific-heat
of  integrable 1-D  Hubbard model under  open-boundary conditions}
\author{Ruihong Yue\thanks{e-mail:
 yue@phys.ocha.ac.jp}
      ~ and~
  Tetsuo Deguchi  \thanks{e-mail: 
deguchi@phys.ocha.ac.jp}   \\[.3cm]
      Department of Physics \\
                  and \\  
Graduate  School  of Humanities and Sciences \\  
      Ochanomizu University \\
                               \\ 
2-1-1 Ohtsuka, Bunkyo-Ku \\ 
     Tokyo 112, Japan}
%\date{\today}
\date{}
\maketitle
\begin{abstract}

The magnetic susceptibility and the  low-temperature specific heat 
of the 1-dimensional Hubbard model under the integrable open-boundary
conditions are  discussed through the Bethe ansatz  with the string
hypothesis. The contributions of the boundary fields to both the
susceptibility and the specific heat are obtained, and their exact
expressions are analytically derived.

\vspace{.5cm}

\end{abstract} 
%\raisebox{15cm}[][]{\hspace{12cm} \bf OCHA-PP-**\\} 

\newpage 

\section{Introduction}

The problem of a quantum impurity in 3D electrons 
has been a fascinating topic in condensed matter physics. 
It is  related to many nontrivial 
phenemena such as   the Kondo problem.  
The low temperature behaviours of the Kondo  and the Anderson models 
are rigorously obtained  by the Bethe ansatz method. \cite{andrei,tsvelick}  
Recently,  some aspects of 
the impurity problem in 1D systems 
have attracted renewd interest from the different branches of physics. 
\cite{Kane,Eggert,SEA,affleckludwig,Furusaki} 
The motivation of this paper is to investigate the  impurity effect 
in the 1-dimensional interacting electrons under integrable boundary
conditions.

Generally, it is expected that 
at low temperature the quantum impurity Hamiltonians 
are renormalized to critical points which correspond to 
conformal invariant boundary conditions. \cite{Eggert,SEA} 
In fact, the impurity effects have been discussed from the 
analysis of the different boundary conditions. 
\cite{affleckludwig,Wongaff,Satsv,Ess}  During the last decade 
the Bethe ansatz techniques for the integrable open chains have been developed.  
\cite{Skl,Mecnep,Forkar,Veggo,Yuefh} 
In Ref. \cite{Satsv}, the magnetization of an anisotropic
Heisenberg model with open-boundary condition was derived. The result is 
generalized  to the supersymmetric t-J model with open
boundaries and the bulk and surface magnetizations  are obtained.  \cite{Ess}  
We can also discuss the Bethe ansatz equations 
for the 1D Hubbard model under some open-boundary conditions. 
\cite{Schulz,Asakawa,Degyue,Shiwad}
However, the thermodynamic properties of  the t-J or Hubbard models  
have not yet been discussed under the open-boundary condition.  
It seems that under the open boundaries the free energy becomes divergent 
due to an infinite number of zero modes.

Let us discuss one of the most different properties of the 1D impurity effect.  
 In the Kondo problem the Wilson ratio  plays an important role. \cite{wilson}   
Based on the local Fermi liquid theory, the local impurity effect in 
the 3 dimensional free electrons is fully characterized by the single parameter  
which is related to the ratio of the specific heat and 
the magnetic susceptibility due to  the impurity.
\cite{nozieres}
For the 1D Hubbard model, however, 
the local impurity effect could not be described by one parameter 
and it should be highly nontrivial. The 
low-energy spectrum  
is given by the Tomonaga-Luttinger 
liquid, which is completely different from the Fermi liquid.  
Furthermore, the impurity effect also depends on 
the Coulomb interaction among the conduction electrons. 
In this paper, we will  evaluate the boundary
effects to the magnetic susceptibility and  specific heat of 
the 1D Hubbard model under the open-boundary conditions, 
which should characterize   
the impurity effect in the 1D interacting electronic system  
in the same way as the Wilson ratio for the Kondo problem.

Let us first review  the Hamiltonian on the 1-dim. lattice with $L$
sites
\cite{Degyue} 
\begin{equation}
\begin{array}{rcl}
{\cal H}
&=& \displaystyle -\sum_{j=1}^{L-1}\sum_{\sigma=\uparrow,\downarrow}
      \left( c_{j\sigma}^{\dagger}c_{j+1\sigma}
       +c_{j+1\sigma}^{\dagger}c_{j\sigma}\right)
       +U\sum_{j=1}^Ln_{j\uparrow}n_{j\downarrow}
       +\mu\sum_{j=1}^L(n_{j\uparrow}+n_{j\downarrow})
       \\[3mm]
& &\displaystyle-\frac{h}2\sum_{j=1}^L(n_{j\uparrow}-n_{j\downarrow})
   +\sum_{\sigma=\uparrow,\downarrow}(p_{1\sigma}n_{1\sigma}
    +p_{L\sigma}n_{L\sigma}).
\end{array}
\label{Ham} 
\end{equation}
Here the symbols $-\mu$ and  $-p_{j\sigma}$ ($j$=1,or $L$) correspond to  
  the chemical potential and boundary fields, respectively. 
The symbol $U$ denotes the Coulomb interaction. When
$p_{j\uparrow}=p_{j\downarrow}$ (boundary chemical potential) or 
$p_{j\uparrow}=-p_{j\downarrow}$ (boundary magnetic field) for $j=$ 1 and
$L$, 
we can solve the above Hamiltonian by the Bethe ansatz method.    
The solution of  $N$ electrons with $M$ down-spins 
has  wave numbers $k_j$ for $j=1, \ldots N$ and  rapidities $v_m$ for 
$m=1 \ldots M$. The Bethe Ansatz equations for it 
are given in the following.  \cite{Degyue}
\begin{eqnarray}
\label{BA1}
\lefteqn{\frac{(e^{-ik_j}p_{1\uparrow}+1)(e^{ik_j}+p_{L\uparrow})}
             {(e^{ik_j}p_{1\uparrow}+1)(e^{-ik_j}+p_{L\uparrow})}
         e^{i2k_jL} } \nonumber \\[3mm]
& &=\displaystyle \prod_{m=1}^M
   \frac{(\sin k_j-v_m+iU/4)(\sin k_j+v_m+iU/4)}
        {(\sin k_j-v_m-iU/4)(\sin k_j+v_m-iU/4)} \quad, 
\end{eqnarray}
\begin{equation}
\label{BA2}
\begin{array}{rcl}
& &\displaystyle\frac{(\zeta_+-v_m-iU/4)(\zeta_--v_m-iU/4)}
             {(\zeta_++v_m-iU/4)(\zeta_-+v_m-iU/4)}
\prod_{n=1, \, n\neq m}^M\frac{(v_m-v_n+iU/2)(v_m+v_n+iU/2)}
{(v_m-v_n-iU/2)(v_m+v_n-iU/2)}  \\[3mm]
& =& \displaystyle\prod_{j=1}^N\frac{(v_m-\sin k_j+iU/4)
     (v_m+\sin k_j+iU/4)}
        {(v_m-\sin k_j-iU/4)(v_m+\sin k_j-iU/4)}  \quad , 
\end{array}
\end{equation}
where 
\begin{equation}
\begin{array}{ccc}
\zeta_+=\left\{\begin{array}{lll}
               \infty & \mbox{for}&p_{1\uparrow}=p_{1\downarrow}\\
               \displaystyle-\frac{1-p_{1\uparrow}^2}{2ip_{1\uparrow}}
               & \mbox{for} & p_{1\uparrow}=-p_{1\uparrow}
               \end{array} \right.  &, &
\zeta_-=\left\{\begin{array}{lll}
               \infty & \mbox{for}&p_{L\uparrow}=p_{L\downarrow}\\
               \displaystyle-\frac{1-p_{L\uparrow}^2}{2ip_{L\uparrow}}
               & \mbox{for} & p_{L\uparrow}=-p_{L\uparrow}
               \end{array} \right. \quad .
\end{array}
\end{equation}
In Ref. \cite{Degyue}, the Bethe ansatz equations (\ref{BA1}) and (\ref{BA2}) 
are systematically derived by using the reflection equations.

\section{The magnetic susceptiblity} 
Now, we discuss the derivation of the magnetic susceptibility 
of the Hamiltonian (\ref{Ham}) 
at zero temperature. We shall evaluate the boundary contributions to the 
suceptibility  using 
the method of  Ref. \cite{Woypen} where the susceptibility is 
derived  under  the periodic boundary 
condition. For the evaluation of the susceptibility 
we assume that the electron density $n=N/L$ is less than half-filling 
($0< n <1$).

Let us consider the ground states of the open-boundary Hubbard model for
the  repulsive ($U>0$) and  attractive ($U<0$ ) cases. 
The equations (\ref{BA1}) and (\ref{BA2}) hold for 
positive and negative values of $U$.    
 However, the  ground-state solutions of the Bethe Ansatz
equations are different for the two cases. 
For $U>0$,  we may consider only the case 
when the ground state is characterized by real $k_j$'s and real $v_m$'s 
\footnote{For some values of the boundary fields,
the Bethe Ansatz equations may have pure imaginary solutions (boundary
bound states) 
  for the ground state.}.  For $U<0$, we may assume that 
the electrons may form singlet bound pairs in the ground state; 
the ground-state solution  of the Bethe Ansatz equations 
consists of real $k_j$'s, real $v_m $'s and  pairs of complex momenta 
$k^{\pm}_n$   
\begin{equation}
\sin(k_n^{\pm})=v_n\pm iu , 
\end{equation}
where $u=|U|/4$.

We solve the Bethe Ansatz equations 
based on the assumptions of the ground state. 
Hereafter, a subscript $r=``>"$ ( $r=``<"$ ) stands for the positive 
(negative) $U$ case.  For the positive $U$ case, let 
$\rho_{>,L}(k)_{1}$ and $\rho_{>,L}(v)_{2}$ denote   
the densities of electron momenta $k_j$ 
and that of down-spin rapidities $v_m$ , respectively.  
 For the negative $U$ case, we denote by $\rho_{<,L}(k)_{1}$ and  
$\rho_{<,L}(v)_2$,  
the densities of real momenta $k_j$ and that of the string centers $v_m$,
respectively.  
\cite{Woypen}
We denote by ${\bm \rho_{r,L}}$  the vector of the densities 
${\bm \rho_{r,L}}=(\rho_{r,L}(k)_{1},\rho_{r,L}(v)_2)$ 
for $r=>,<$. We now take the asymptotic expansion with respect to $1/L$. 
Then, we have the following integral equation  
\begin{eqnarray}
{\bm \rho}_{r,L}(k,v)
&=&\displaystyle 
{\bm \rho}^0_{r,L}(k,v)
   +{\bf K}_r(k,v|k',v'){\bm \rho}_{r,L}(k',v')  .     
\label{density}
\end{eqnarray}
Here the initial values of the densities ${\bm \rho}^0_{r,L}(k,v)= 
{\bm \rho}^0_{r, \infty}(k,v)+{\bm \tau}^0_r/L$ are given by  
\begin{equation}
{\bm \rho}^0_{>,\infty}(k,v)=     
\left(\begin{array}{c}  
\displaystyle \frac1{\pi}\\[3mm]
               0 
\end{array}\right) \qquad ,\qquad
{\bm \tau}^0_{>}(k,v)=\displaystyle 
\left(\begin{array}{c} 
                {\frac d {dk} } P_{>,0}(k)\\[3mm]
                 {\frac d{dv}} Q_{>,0}(v)
      \end{array}\right)
\end{equation}
\begin{equation}
{\bm \rho}^0_{<,\infty}(k,v)=     
\left(\begin{array}{c}  
\displaystyle{\frac1{\pi}}\\[3mm]                
\displaystyle\frac2{\pi}Re\frac1{\sqrt{1-(v-iu)^2}}
     \end{array}\right)~~,~~
{\bm \tau}^0_{<}(k,v)=\displaystyle \left(\begin{array}{c} 
                {\frac d{dk}} P_{<,0}(k)\\[3mm]
                 {\frac d{dv}}Q_{<,0}(v)
      \end{array}\right) . 
\end{equation}
The definitions of 
$d/dk P_{r,0}$(k) and $d/dv Q_{r,0}(v)$ will be given  in eq. (\ref{BAE})
 for $r=>$ 
\footnote{$d/dv Q_{>,0}(v)$ is given by $d/dv Q_{0}^n(v)$ of $n=1$ in eq.
(\ref{BAE}). }, 
while   for $r=<$ they are given in the following
\begin{equation}
\begin{array}{rcl}
2 \pi {\frac d {dk}} P_{<,0}(k)&=&\displaystyle 
    \frac{2(1+p_{L\uparrow}\cos k)}
               {1+p_{L\uparrow}^2+2p_{L\uparrow}\cos k}
       - \frac{2p_{1\uparrow}(p_{1\uparrow}+\cos k)}
                            {1+p_{1\uparrow}^2+2p_{1\uparrow}\cos k}
       + \frac{2u \cos k}{\sin^2 k+u^2}\\[6mm]
2 \pi {\frac d {dv}} Q_{<,0}(v)&=&\displaystyle 
         Re\left(\frac{1+p_{L\uparrow}\sqrt{1-(v-iu)^2}}
                       {1+p_{L\uparrow}^2+2p_{L\uparrow}\sqrt{1-(v-iu)^2}}
         \frac2{\sqrt{1-(v-iu)^2}}\right)\\[6mm]
 & &\displaystyle - Re\left(
        \frac{p_{1\uparrow}+\sqrt{1-(v-iu)^2}}
         {1+p_{1\uparrow}^2+2p_{1\uparrow}\sqrt{1-(v-iu)^2}}
         \frac{2p_{1\uparrow}}{\sqrt{1-(v-iu)^2}}\right)\\[6mm]
& &\displaystyle
   + \frac{4u}{v^2+4u^2}
  +\frac{2(u+i\zeta_+)}{v^2+(u+i\zeta_+)^2}+
\frac{2(u+i\zeta_-)}{v^2+(u+i\zeta_-)^2}\\[3mm]
\end{array}
\end{equation}
We note that the kernel ${\bm K}_{>}$ was given in \cite{Degyue} and  
${\bm K}_{<}=\sigma^3{\bm K}_{>}\sigma^3$ \cite{Woypen}.  
The parameters $Q$ and $B$ \cite{Degyue} 
for the upper- or lower-bounds of the integral
intervals have 
the following constraints 
\begin{equation}
\label{constraint}
\begin{array}{ccc}
\displaystyle \int_{-Q}^Q\rho_{>,L}(k)_1dk=2n+\frac{1}{L}&,&
\displaystyle \int_{-B}^B\rho_{>,L}(v)_2dv=n-2s+\frac{1}{L}\\[3mm]
\displaystyle \int_{-Q}^Q\rho_{<,L}(k)_1dk=4s+\frac{1}{L}&,&
\displaystyle \int_{-B}^B\rho_{<,L}(v)_2dv=n-2s+\frac{1}{L}
\end{array}
\end{equation}
where $s$ has been defined by $s=(N-2M)/2L$. 
The ground-state energy $E_r$ for $r=>,<$ is  given by
\begin{equation}
\begin{array}{rcl}
\displaystyle{\frac {E_{>}} L}&=&\displaystyle\frac1L[1-\mu_s-h/2]+({\bm
e}^0_{>},{\bm
           \rho}_{>,L})\\[3mm]
\displaystyle{\frac {E_{<}}
L}&=&\displaystyle\frac1L[1-\mu_s-\mu+2\sqrt{1+u^2}]+
          ({\bm e}^0_{<},{\bm \rho}_{<,L})
\end{array}
\end{equation}
where $\mu_s=\mu/2-h/4$. The dressed energy ${\bm e}_r$ satisfies 
${\bm e}_r = {\bm e}_r^0 + {\bm K}^T_r {\bm e}_r$ with  the initial values 
\begin{equation}
{\bm e}^0_{>}=\left(\begin{array}{c}\mu_s-\cos k \\ h/2
                       \end{array}\right)~~,
{\bm e}^0_{<}=\left(\begin{array}{c}\mu_s-\cos k \\ 
                 -2Re\sqrt{1-(v-iu)^2} \end{array}\right)
\end{equation}

By minimizing  the ground state energy with respect 
to the variable $s$,  we have  the following functional relation between
$s$ and $h$ through  
the constraints  (\ref{constraint}). 
\begin{equation}
\label{mag-fields}
h =\displaystyle 2 \frac{\epsilon_{r}(Q)_1\zeta_{r}(B)_2-
         \epsilon_{r}(B)_2\zeta_{r}(Q)_1}{\det
\xi_{r}(Q,B)} \qquad {\rm for } \qquad r = >, < . 
\end{equation}
The dressed charge matrix ${\bm \xi}_r$ is defined by the relation 
${\bm \xi}_{r}(k,v)={\bf 1}+{\bm K}^{T}_r(k,v|k',v')
{\bm \xi}_{r}(k',v')$,  and the symbols $\zeta_{r,j}$ are given by the
matrix elements of
the dressed charge ${\bm \xi}_r$ as follows;  
when $r=>$ $\zeta_{>,j}=(\xi_{>})_{j1}$ for $j=1,2$, and  
when $r=<$ $\zeta_{<,j}=(\xi_{<})_{j1}+2(\xi_{<})_{j2}$ 
for $j=1,2$. The symbols 
${\bm \epsilon}_r$ are defined by 
\begin{equation}
{\bm \epsilon}^0_{>}= {\bm e}_{>}^0 + 
\left(\begin{array}{c}
 h/4 \\ -h/2 
\end{array}\right)
\qquad ,\qquad 
{\bm \epsilon}^0_{<}= {\bm e}_{<}^0 + 
\left(\begin{array}{c}
 h/4 \\  0 
\end{array}\right)
\end{equation}

We can calculate the magnetization $s$ 
through the equations (\ref{mag-fields}) and 
(\ref{constraint}).
We thus derive  the magnetic susceptibility $\chi_{r,L}$ 
of the finite-lattice of $L$ sites, for the $r=>$ and $r=<$ cases. 
Here $L$ is  a large but finite number. 
\begin{equation}
\begin{array}{rcl}
\chi_{r,L} &=&\displaystyle
        \left\{\frac{\partial h}{\partial Q}\frac{\partial Q}{\partial S}
         +\frac{\partial h}{\partial B}\frac{\partial B}{\partial S}
         \right\}^{-1}\\[3mm] 
&=&\displaystyle \left\{\frac{2 v_{r,1}(Q)}{\rho_{r,L}(Q)_1}
       \frac{\zeta^2_{r,2}(B)}{\det^2 {\bm \xi}_r(Q,B)}+
 \frac{2 v_{r,2}(B)}{\rho_{r,L}(B)_2}\frac{\zeta^2_{r,1}(Q)}
      {\det^2 {\bm \xi}_r(Q,B)}\right\}^{-1}
\end{array} \qquad {\rm for } \qquad r = >, <  . 
\label{suscept}
\end{equation}
The Fermi velocity $v_{r,j} $ are given  by the $j$-th component of 
${\bm v}_r$ defined by ${\bm v}_r= {\bm e}_{r}^{'0} +{\bm K}_r{\bm v}_r$, 
where ${\bm e}_{r}^{'0}$ denote the vector whose first and second
components 
are  $d/dk {e}_r(k)_{1}$ and $d/dv e_r(v)_2$, respectively. 
We note that if we specify the density $n$ and magnetization $s$, then
the parameters $Q$ and $B$ in eq. (\ref{suscept}) are defined by
(\ref{constraint}).

Let us discuss the boundary contribution $\delta \chi_r$ to the
susceptibility for both 
the repuslive and attractive cases. We assume that  
 the finite and the infinite systems have  the same $n$ and $s$.   
Then we may formally define $\delta \chi_r$ by the following  
\begin{equation} 
\delta \chi_r = \chi_{r,L}(Q,B) - \chi_{r,\infty}(Q_{\infty},B_{\infty}),  
\qquad {\rm for } \qquad r = >, < . 
\end{equation} 
Here $Q_{\infty}$ and $B_{\infty}$ are the interval parameters for the
infinite system, 
which are given by eq. (\ref{constraint}) after taking the infinite limit: 
$L \rightarrow \infty$.  
For the case of nonzero magnetic field ($B_{\infty} \ne \infty$), we may
evaluate 
$\delta B= B-B_{\infty}$ by taking the derivatives of eq.
(\ref{constraint}). 
In terms of the dressed charge we have 
\begin{equation} 
\left(
\begin{array}{c}
\delta Q \\ 
\delta B
\end{array} \right)
= \displaystyle  {\frac 1L}
\left(
\begin{array}{cc} 
\xi_{r,22} & -\xi_{r,21} \\  
-\xi_{r,12} & \xi_{r,11} 
\end{array} \right) 
\left(
\begin{array}{c}  
1- \displaystyle \int_{-Q_{\infty}}^{Q_{\infty}} \tau_{r}(k)_1 dk    \\ 
1- \displaystyle \int_{-B_{\infty}}^{B_{\infty}} \tau_{r}(v)_2 dv    
\end{array} \right) {\frac 1 {\det{\bm \xi}_{r}} } \qquad {\rm for } \qquad
r = >, < . 
\end{equation}
Here the matrix elemnts of the dressed charge are evaluated at $Q_{\infty}$
and $B_{\infty}$. 
For $U>0$, by the Wiener-Hopf method we can show that under zero magnetic
field 
$B$ is as large as $\log L$. 
We also note that for  some values of the boundary fields,  
the magnetization $s$ of the finite system can  take a nonzero value 
under zero magnetic field: $h$=0, in general.

We find that 
the boundary contribution $\delta \chi_r$ 
to the magnetic suceptibility contains both  the charge and spin parts.
We recall that the densities of the rapidities contain the 
$1/L$-terms, which come  from the open-boundary condidtion. 
We note that under the periodic boundary condition, the finite-size
corrections of the densities do not have $1/L$-terms; the first nonzero order is given
by $1/L^2$-terms.   
Therefore, the $1/L$-term together with $\delta Q$ and $\delta B$ will 
reflect the effect of the 
open-boundary conditions. 
It seems that the result is  different from the perturbative 
calculation of the $\delta \chi$ in Ref. \cite{Furusaki}. However, 
the  $\delta \chi$ in Ref. \cite{Furusaki} is obtained by using the  
bosonization method, where some limiting procedures are employed. 
Thus, it is not easy to point out  the most important reason why they are different. 
We shall discuss   this possible discrepancy in later publications. 

\section{The specific heat}
In the rest of the paper, we show how to calculate  the
low-temperature specific heat 
of the open-boundary Hubbard model.  We consider only the repulsive case 
($U>0$).  For the negative $U$ case, we can derive the 
similar results making use of the particle-hole transformation.
Under the zero boundary fields case, the Bethe ansatz equations (2) and (3)
are equivalent to  those of the periodic case. Thus, 
the solutions of the Bethe ansatz equations have the same structure
such as in the periodic Hubbard model. \cite{Tak} There are both real and complex
solutions for the momentum $k_j$ and rapility $ v_m$.
They can be classified into 
three groups:  real momenta $k$, 
 n-$\lambda$ strings and n-$\lambda$-k strings. The n-$\lambda$ string solution for rapility
$v$ is given by 
\begin{equation}
\lambda_m^{n,j}=\lambda_m+iu(n+1-2j), \qquad j=1,\cdots, n .\\[3mm]
\end{equation}
The n-$\lambda$-k string solutions for the momentum $k$ and the rapility $v$  are defined by
\begin{equation}
\begin{array}{rcl}
\lambda_m^{'n,j}&=&\displaystyle 
     \lambda'_m+iu(n+1-2j), \qquad j=1,\cdots, n \\[3mm]
k^{n,2j+1}_m&=&\displaystyle 
         \pi-\sin^{-1}(\lambda'_m+iu(n-2j)), \qquad 0\leq j\leq n-1\\[3mm]
k^{n,2j}_m&=&\displaystyle 
        \sin^{-1}(\lambda'_m+iu(n-2j)), \qquad 1\leq j\leq n-1,\\[3mm] 
k^{n,2n}_m&=&\displaystyle \pi-\sin^{-1}(\lambda'_m-iun).
\end{array}
\label{string} 
\end{equation}
Here  $\lambda_n$ and $\lambda'_n$ are the centers  of
an $n-\lambda$ string 
and an $n-\lambda-k$ string, respectively. 
The symbols $M_n$,  $M'_n$ and $M'$  are the numbers of 
 n-$\lambda$ strings, n-$\lambda$-k strings and the all $\lambda$-k 
strings, respectively.  The $k_j$'s form real momenta for $j=1,\cdots, N-2M'$.
We note $M'= \sum_n nM'_n$.  We recall $u=U/4$.  In this paper, we only  consider 
the case where all solutions are given by the three groups with the strings.

Taking the asymptotic expansion ( up to the order of $1/L$ ), we can derive
the following integral equations of the  densities of particles and holes 
from the Bethe Ansatz equations (\ref{BA1}) and (\ref{BA2}).  
\begin{equation} 
\begin{array}{rcl}
\rho^h(k)
&=& \displaystyle - \rho(k)+ \frac1{\pi}+\frac1{L}\frac{d}{dk}P_0(k) 
   +\sum_{n=1}^{\infty}\int\theta'_{n}(\sin k-\lambda)\cos k
     \left(\sigma_n(\lambda)+\tilde{\sigma}_n
     (\lambda)\right )d\lambda\\[5mm]
\sigma^h_n(\lambda)
&=&\displaystyle \frac{1}{L}\frac{d}{d\lambda}Q_0^n(\lambda)
    + \int\theta'_{n}(\sin k-\lambda)\rho(k)dk 
    -\sum_{m=1}^{\infty}A_{nm} \sigma_m(\lambda)\\[5mm]
\tilde{\sigma}^h_n(\lambda)
&=&\displaystyle 
\frac2{\pi}Re\left(\frac1{\sqrt{1-(\lambda-inu)^2}}\right)
   +\frac1L\frac{d}{d\lambda}\tilde{Q}_0^n(\lambda) \\ 
&  & + \displaystyle 
 \int\theta'_{n}(\sin k-\lambda)\rho(k)dk
        -\sum_{m=1}^{\infty}A_{nm}\tilde{\sigma}_m(\lambda)\\[5mm]
P_0(k)
&=&\displaystyle \frac1{2\pi i}\log
         \frac{(1+p_{1\uparrow}e^{-ik})(p_{L\uparrow}+e^{ik})}
         {(1+p_{1\uparrow}e^{ik})(p_{L\uparrow}+e^{-ik})} \\ 
        & & \displaystyle -\sum_{m=1}^{\infty} 
          \theta_{m}(\sin k)(2-\delta(M_m,0)-\delta(M_m^{'},0))
\\[5mm]
Q_0^n(\lambda)
&=&\displaystyle \sum_{m=1}^{\infty}\Theta_{nm}
    (\lambda)(1-\delta(M_m,0)) \\
  & & - \displaystyle {\frac1{2\pi i}}\sum_{j=1}^{n} \log
     \frac{(\lambda+\zeta_++iu(n-2j))(\lambda+\zeta_-+iu(n-2j))}
          {(\lambda-\zeta_+-iu(n-2j))(\lambda+\zeta_--iu(n-2j))}
\\[5mm]
\tilde{Q}^n_0(\lambda)
&=&\displaystyle \sum_{m=1}^{\infty}  \Theta_{nm}(\lambda)
(1-\delta(M_m^{'},0)) \\ 
& & \displaystyle  -\frac1{2\pi i}\sum_{j=1}^{2n}\log 
   \frac{(1+p_{1\uparrow}e^{-ik^{n,j}})(p_{L\uparrow}+e^{ik^{n,j}})}
       {(1+p_{1\uparrow}e^{ik^{n,j}})(p_{L\uparrow}+e^{-ik^{n,j}})}
\\[5mm]
& &\displaystyle+\frac1{2\pi i}\sum_{j=1}^{n}\log
   \frac{(\lambda+\zeta_++iu(n-2j))(\lambda+\zeta_-+iu(n-2j))}
        {(\lambda-\zeta_+-iu(n-2j))(\lambda+\zeta_--iu(n-2j))}
\\[5mm]
A_{nm}f(x)&= &\displaystyle  \delta_{nm}f(x)+{\frac {\partial} {\partial
x}}  \int\Theta_{nm}(x-x')f(x')dx'
\end{array}
\label{BAE} 
\end{equation}
where 
$\theta_n(x)=2\tan^{-1}(x/nu)/2\pi$ and $\Theta_{nm}(x)
=(1-\delta_{nm})\theta_{|n-m|}(x) + 2\theta_{|n-m|+2}(x)+
\cdots +2\theta_{n+m-2}(x) +\theta_{n+m}(x)$.  
The symbol $\delta(j,k)$ denotes the Kronecker delta. 
We note that $\rho, \sigma_n$ and ${\tilde \sigma}_n$ are 
the particle densities of the real momenta $k_j$'s, the centers of the 
$n-\Lambda$ strings, and the centers of the $n-\Lambda-k$ strings,
respectively;  
$\rho^h, \sigma_n^h$ and ${\tilde \sigma}_n^h$ are the hole densities 
of them, respectively. It is remarked that 
in the derivation of the Bethe ansatz equations (\ref{BAE}), we have
assumed that 
the string solutions for the finite system could have small deviations 
from those of the inifnite system given in (\ref{string}).

The total energy $E$ of the system is given by 
\begin{equation}
\begin{array}{rcl}
\displaystyle \frac{E}{L}&=&\displaystyle
\frac1{L} \left( (1-\mu_s)(1-\delta(N,2M^{'})) 
 - \frac{h}2 \sum_{n=1}^{\infty}n(1-\delta(M_n,0)) \right) \\ 
& & \displaystyle - \frac 1L  \sum_{n=1}^{\infty}(2\sqrt{1+(nu)^2}+ n \mu)
(1-\delta(M'_n,0))  \\[5mm]
&&\displaystyle +\int_{-\pi}^{\pi}(\mu_s-\cos k)\rho(k)dk
  +\frac{h}{2}\sum_{n=1}^{\infty}n \int_{-\infty}^{\infty}
\sigma_n(\lambda)d\lambda\\[5mm]
&&\displaystyle  + \sum_{n=1}^{\infty}\int_{-\infty}^{\infty}
2Re\left(\sqrt{1-(\lambda-inu)^2}+n \mu \right)
\tilde{\sigma}_n(\lambda)d\lambda
\end{array} . 
\label{energy} 
\end{equation}
It should be emphasized that the sums for the zero modes in (\ref{energy}) 
do not become infinite because we take $L$ a large but finite number.

Minimizing the thermodynamic potential 
$\Omega=E-TS$, we can show  the following 
thermal Bethe Ansatz equations for the ratios of the particle and hole
densities 
$\zeta=\rho^h/\rho$,  $\eta_n=\sigma^h_n/\sigma_n$, and  
$\tilde{\eta}_n=\tilde{\sigma}^h_n/\tilde{\sigma}_n$ 
\begin{equation}
\begin{array}{rcl}
\ln \zeta(k)
&=&\displaystyle -\frac{2 \cos k}{T}+\int_{-\infty}^{\infty}
   \mbox{sech}(\frac{\pi(\lambda-\sin k)}{2u})\left(
    \ln\frac{1+\tilde{\eta}_1}{1+\eta_1}
    -\frac{4Re(\sqrt{1-(\lambda-iu)^2}}{T}\right)
   \frac{d\lambda}{4u}\\[5mm]
\ln\eta_1(\lambda)
&=&\displaystyle s*\left(\ln(1+\eta_2(\lambda))
   -\int_{-\pi}^{\pi}\ln(1+\zeta^{-1}(k)\cos k \, \delta(\lambda
   -\sin k)dk\right)\\[5mm]
\ln\tilde{\eta}_1(\lambda)
&=&\displaystyle s*\left(\ln(1+\tilde{\eta}_2(\lambda))
    -\int_{-\pi}^{\pi}\ln(1+\zeta(k)\cos k \, \delta(\lambda
    -\sin k)dk\right)\\[5mm]
\ln\eta_n(\lambda)
&=&\displaystyle s*\left(\ln(1+\eta_{n+1}(\lambda))+
   \ln(1+\zeta_{n-1}(\lambda))\right)\\[5mm]
\ln\tilde{\eta}_n(\lambda)
&=&\displaystyle s*\left(\ln(1+\tilde{\eta}_{n+1}(\lambda))+
   \ln(1+\tilde{\zeta}_{n-1}(\lambda))\right)\\[5mm]
\ln\eta_n(\lambda)
&\stackrel{n\rightarrow \infty}{=}&\displaystyle n\frac{h}{T}\\[5mm]
\ln\tilde{\eta}_n(\lambda)
&\stackrel{n\rightarrow \infty}{=}&\displaystyle n\frac{4u+2\mu}{T}
\end{array}
\label{hole-eq} 
\end{equation}
where the convolution $s*$ is given by  $s*f(x)=\int(
\mbox{sech}(\pi(x-x')/2u)/4u)f(x')dx'$. 
The last two limits in (\ref{hole-eq}) should be understood  
as the asymptotic limit of $n$ consistent that of $1/L$; the notation $n
\rightarrow \infty$ 
means that $n$ should be taken a very large but finite number.  
We should note that the variation of the particle and hole densities
should be taken over all the positive values of rapidities since they are 
even functions; we have such as $\rho(k) =  \rho(-k)$, and so on:    
$\sigma(v) =  \sigma(-v)$, and  
${\tilde \sigma}(v) = {\tilde \sigma}(-v)$, $\rho^h(k) =  \rho^h(-k)$,  
$\sigma^h(v) =  \sigma^h(-v)$, and  ${\tilde \sigma}^h(v) = {\tilde
\sigma}^h(-v)$.

Substituting (\ref{BAE}) into $\Omega$ we have the asymptotic expansion of 
the thermodynamic potential $\omega_L=\Omega/L$  with respect to $1/L$
\begin{equation}
\begin{array}{rcl}
\omega_L &=&\displaystyle
          \frac1{L}\left((1-\mu_s) (1-\delta(N, 2M'))
-\frac{h}2\sum_{n=1}^{\infty}
          (1-\delta(M_n,0)) \right) \\ 
& & - \displaystyle {\frac 1L} \sum_{n=1}^{\infty}(2\sqrt{1+(nu)^2}+n \mu)
          (1-\delta(M'_n,0))\\[5mm]
& &\displaystyle -\int^{\pi}_{-\pi}\left(\frac{1}{2\pi}
   +{\frac 1{2L}}
\frac{dP_0(k)}{dk}\right)T\ln (1+\zeta^{-1}(k))dk
   -\sum_{n=1}^{\infty}\int_{-\infty}^{\infty}
{\frac 1 {2L}}   
\frac{dQ^n_0(\lambda)}{d\lambda}
   T\ln(1+\eta^{-1}_n(\lambda)d\lambda\\[5mm]
&&\displaystyle -\sum_{n=1}^{\infty}\int_{-\infty}^{\infty}
\left( {\frac1{\pi}}Re(\frac1{\sqrt{1-(\lambda-inu)^2}}
  +{\frac 1{2L}}
\frac{d\tilde{Q}^n_0(\lambda)}{d\lambda}  \right)
   T\ln(1+\tilde{\eta}^{-1}_n)d\lambda +o(\frac{1}{L})
\end{array}
\label{TBA} 
\end{equation}

We now introduce $\kappa=T\ln\zeta$ and $\epsilon_1=T\ln \eta_1$.  
We denote the zero-temperature limits of $\kappa$ and $\epsilon_1$
by $\kappa^{(0)}$ and $\epsilon_1^{(0)}$, respectively. 
Hereafter we assume $h \gg T$. 
We can calculate the specific heat $C_{L}$ through 
$C_{L}=-T\partial^2 \omega_L/\partial T^2$. 
We give the final results:

\par \noindent 
{\bf a:} \quad {\boldmath $-\mu \leq -2 -h/2$}

\begin{equation}
\omega_L(T,h,\mu)
=\omega_L(0,h,\mu)-\frac{T^{3/2}}{\pi}(1+\frac{\delta_f}{L})
\int_0^{\infty}\ln\left(1+e^{-x^2}\exp\{\frac{2+h/2-\mu}{T}\}\right)dx
\label{free-a} 
\end{equation}
where 
$$\delta_f=\left\{\begin{array}{cc}
                      1/(1+p_{L\uparrow}) &\quad , \quad  p_{L\uparrow}\neq -1 \\
                      1/2                 &\quad , \quad  p_{L\uparrow}=-1
                     \end{array}\right\}-
\left\{\begin{array}{cc} 
     p_{1\uparrow}/(1+p_{1\uparrow})&\quad , \quad  p_{1\uparrow}\neq -1 \\
             1/2                    &\quad , \quad  p_{1\uparrow}=-1
                     \end{array}\right\} . 
$$
Let $C_{\infty}$ and $\delta C$ denote the bulk and the boundary
specific-heats, 
respectively. Then from (\ref{free-a}), we see the following  
\begin{equation}
C_{L}= C_{\infty}+ \delta C \\[5mm]
=\displaystyle C_{\infty} (1+\frac1{L}\delta_f) . 
\end{equation}

\par \noindent 
{\bf b :} \quad {\boldmath $\epsilon_1^{(0)}(0)\geq 0$, \quad $-\mu > -2
-h/2$}

In this region, the thermodynamic potential is given by 
\begin{equation}
\begin{array}{rcl}
\omega_L(T,h,\mu)
&=& \displaystyle \omega_L(0,h,\mu)-\frac{\pi^2T^2}{12\sin Q}
    \left(\frac1{\pi}+\frac1L\frac{d}{dk}P_0(Q)\right)\\[3mm]
& &\displaystyle- 2 T^{3/2}\left[g(0)+\frac1{L}\delta g(0)\right]
    \sqrt{\frac{2}{\frac{d^2}{d\lambda^2}\epsilon_1^{(0)}(0)}}
\int_0^{\infty}\ln\left(1+e^{-x^2}e^{-\epsilon_1^{(0)}(0)/T}
   \right)    dx\\[5mm]
g(\lambda)&=&   \displaystyle {\frac 1 {2\pi}}
\int_{-Q}^{Q}\frac{2u}{(\lambda-\sin
             k)^2+u^2}\frac{dk}{2\pi}\\[5mm]
\delta g(\lambda)&=&   \displaystyle {\frac 1 2} \int_{-Q}^{Q}\frac{2u}
{(\lambda-\sin k)^2+u^2} \frac{d}{dk} P_0(k)\frac{dk}{2\pi}
         + {\frac 1 2 } {\frac{d}{d\lambda}} Q_0^1(\lambda)
\end{array}
\end{equation}
Here the paramter $Q$ is the zero of $\kappa^{(0)}(k)$. 
 From this expression, we  find  the ratio 
$\delta C/ C_{\infty} $ 
\begin{equation}
\frac{\delta C}{ C_{\infty}}= \frac{\pi}{L}\frac{d}{dk}P_0(Q)
\end{equation}

\par \noindent 
{\bf c:}\quad {\boldmath $h\geq 4(\sqrt{1+u^2}-u)$, \quad $-\mu \geq 2-h/2$
}

\par 
In this region, we find 
\begin{equation}
\begin{array}{rcl}
\omega_L(T,h,\mu)&=& \displaystyle \omega_L(0,h,\mu)-
               T^{3/2}\pi^{-1}(1+\frac{\pi \frac{d}{dk}P_0(\pi)}{L})
               \int_0^{\infty}\ln(1+e^{\alpha}e^{-x^2})dx\\[5mm]
    & & \displaystyle -
               T^{3/2}4(1+u^2)^{1/4}\pi^{-1}
          (1+{\frac {\Gamma} L}
)\int_0^{\infty}\ln(1+e^{\beta}e^{-x^2})dx\\[5mm]
 \alpha &=& \displaystyle \frac{2+u-h/2}{T}~~,~~
 \beta =  -\frac{h-4(\sqrt{1+u^2}-u)}{T} \\[5mm]
\Gamma &=&\displaystyle
\frac{\pi \sqrt{1+u^2}}{2}\left[ \int_{-\pi}^{\pi}
          \frac{2u}{\sin^2 k+u^2}\frac{{\frac d {dk}} P_0(k)}{2\pi}dk 
+ {\frac d {d \lambda}} Q_0^1(0) \right]
\end{array}
\end{equation}

\par \noindent 
{\bf d:} \quad {\boldmath $\epsilon_1^{(0)}(0)< 0$, \quad
$\kappa^{(0)}(\pi)>0$}

Let us denote by $B$ the zero of $\epsilon_1^{(0)}(\lambda)$.  
Then from the thermal Bethe-Ansatz equations (\ref{TBA}), we obtain  
\begin{equation}
\omega_L(T,h,\mu)= \omega_L(0,h,\mu)
   -\frac{\pi^2T^2}{6{\frac d {dk}} \kappa^{(0)}(Q)}\rho_L^c(Q)
   -\frac{\pi^2T^2}{6{\frac d {d\lambda}}\epsilon^{(0)}_1(B)}\rho_L^s(B)
\end{equation}
where the density functions are given by 
$\rho_L^{c}(k)=\rho_{>,L}(k)_1$ and $\rho_L^{s}(v)=\rho_{>,L}(v)_2$, where 
${\bm \rho}_{>,L}=(\rho_{>,L}(k)_1,\rho_{>,L}(v)_2)$ 
is the density for the ground state of the repulsive case.   
(See also  Ref. \cite{Degyue}). 
\vskip 0.6cm 

\par \noindent 
{\bf e:} \quad {\boldmath $4\sqrt{1+u^2}-u > h \gg T$,\quad 
$\kappa^{(0)}(\pi) \leq 0$ }

In this region, the free energy can be evaluated  as 
\begin{equation}
\begin{array}{rcl}
\omega_L(T,\mu,h)&=& \displaystyle \omega_L(0,\mu,h)
 -\frac{\pi^2T^2}{6{\frac d {d\lambda}} \epsilon^{(0)}_1(B)}
\sigma_1^{(0)}(B)\\[3mm]
&& \displaystyle - T^{3/2}\rho_0(\pi)\sqrt{ \frac {2}{-{\frac {d^2}
{dk^2}} \kappa^{(0)}(\pi)}   }
 \int^{\infty}_0\ln (1+e^{\kappa^{(0)}(\pi)/T} e^{-x^2}  )dx
\end{array}
\end{equation} 
where $\rho_0 $ and $\sigma_1^{(0)}$ are defined by
\begin{equation}
\begin{array}{rcl}
\sigma_1^{(0)}(\lambda)
&=&\displaystyle \sigma_0(\lambda)+
    \int_{|\lambda'|>B}R(\lambda-\lambda')\sigma_1^{(0)}
    (\lambda')d\lambda\\[3mm]
\frac{d}{d\lambda}\epsilon^{(0)}_1(\lambda)
&=&\displaystyle \int_{-\pi}^{\pi}s(\sin k-\lambda)\cos k dk
   +\int_{|\lambda'|>B} R(\lambda-\lambda')
   \frac{d}{d\lambda'}\epsilon^{(0)}_1(\lambda')d\lambda'\\[3mm]
\sigma_0(\lambda)
&=&\displaystyle \int^{\pi}_{-\pi} \frac1\pi s(\lambda -\sin k) dk 
  +\frac1L \left(\frac{d}{d\lambda} Q^1_0(\lambda)+
    R*\frac{d}{d\lambda}Q^1_0(\lambda) \right)  \\ 
  & & +\displaystyle \frac1L  \int^{\pi}_{-\pi} s(\lambda -\sin k)
\left(\frac{d}{dk} P_0(k)-\hat{Q}_0(k)\right)d k
   \\[3mm]
\rho_0(k)
&=&\displaystyle \frac1\pi +\frac1L\left(\frac{d}{dk}P_0(k)
      -\hat{Q}_0(k)\right)\\[3mm]
& &\displaystyle   +\cos k\int^{\infty}_{-\infty}a_1(\sin k-\lambda)
   \left(\sigma_0(\lambda)
-\frac1L\left[\frac{d}{d\lambda}Q^1_0(\lambda)+R*
   \frac{d}{d\lambda}Q^1_0(\lambda)\right]\right) d\lambda
\end{array}
\end{equation} 
and $\hat{Q}(k) =\cos k\int s(\sin k-x) d/dx Q^1_0(x)dx$. 
Here $f*g(x)=\int f(x-x')g(x')dx'$ , $ s(x)=\mbox{sech}(\pi x/2u)/4u$
and $ R(x)=s*2u/(2\pi(x^2+u^2))$.

 From the low-temperature expansion of the thermodynamic potential, 
we find the boundary contribution to the specific heat at 
 low temperature 
\begin{equation}
\frac{\delta C}{C_{\infty}}=\frac{1}{L}\left\{\begin{array}{ll}
  \delta_f&, \quad \mbox{case a}\\[3mm]
  \pi\frac{d}{dk}P_0(Q) &,\quad \mbox{ case b}\\[3mm]
  \displaystyle \frac{\pi \frac{d}{dk}P_0(\pi)C^c + \Gamma C^s}
                {C^c + C^s} &, \quad \mbox{case c}\\[5mm]
  \displaystyle \left( {\frac{C^c\tau^c(Q)}{\rho^c_{\infty}(Q)}  
                  +\frac{C^s\tau^s(B)}{\rho^s_{\infty}(B)} }\right)/(C^c +
C^s) & ,
\quad \mbox{case d}\\[5mm]
  \displaystyle \frac{\delta\sigma_1^{(0)}(B)}{\sigma^{(0)}_{1,\infty}(B)}
   &, \quad \mbox{case e} 
\end{array}\right. 
\end{equation}
Here $C_{\infty}$ denotes the bulk specific-heat, and   
$C^c$ and  $C^s$ denote the contributions 
to the bulk specific-heat $C_{\infty}$  from the 
charge and spin parts, respectively; 
$C_{\infty} = C^c + C^s$. The symbols 
$\sigma^{(0)}_{1,\infty}(B)$ and $\delta\sigma_1^{(0)}(B)$ are 
the bulk  and the $1/L$ parts of $\sigma_1^{(0)}(B)$.

The boundary contributions to the magnetic susceptibility 
and the  specific heat  depend on  the 
boundary fields and  the electron density. 
In regions $a$, and $c$, the specific heat is proportional to 
$T^{-3/2} e^{-\alpha/T}$ where $\alpha$ is a positive constant. 
In regions $ b$,  $d$ 
and  $e$, the specific heat  depends linearly on temperature. 
For the boundaries of the regions between $a$ 
and $b$, $b$ and $c$, $b$ and $d$,
 $c$ and $e$, and $d$ and $e$, 
the specific heat is proportional to $T^{1/2}$. 
We recall that for some regions of the
boundary fields, there may exist other types of solutions 
(boundary string states)  
of the Bethe ansatz equations. \cite{Skosal,Kapsko,Tsuyam,Bedfra,Essfra}. 
We can calculate the contribution from the boundary string
states simply by 
modifying the term ${\bm \tau}^0_r$ in our derivation.

In summary, we have studied the boundary contributions to the 
magnetic susceptbility and the specific
heat for the 1-dimensional Hubbard model 
under the general open-boundary conditions. 
They are calculated analytically and explicitly. From 
the results  we can discuss exactly 
the impurity effect in the 1-dimensional Hubbard model 
or in the interacting electrons in 1D.

We are  grateful to Dr. K. Kusakaba for helpful discussion on 
the boundary bound states.
R. Yue was granted by the JSPS foundation 
and the Monbusho Grant-in-Aid of Japanese Government.


\begin{thebibliography}{[99]}
\bibitem{andrei} N. Andrei, K. Furuya and J. H. Lowenstein,
         1983 Rev. Mod. Phys. {\bf 55} 331
\bibitem{tsvelick} A.M. Tsvelik and P.B.. Wiegmann, 1983 Adv. in Phys. {\bf 32}  
453  
\bibitem{Kane} C.L. Kane and M.P.A. Fisher, 1992 
       Phys. Rev. Lett. {\bf 68} 1220 
\bibitem{Eggert} S. Eggert and I. Affleck, 1992 Phys. Rev. B {\bf 46}  10866 
\bibitem{SEA} E.S. S$\phi$rensen, S. Eggert and I. Affleck, 1993 J. Phys. 
             {\bf A26} 6757 
\bibitem{affleckludwig}I. Affleck and A. W. W. Ludwig, 
        1992 Phys. Rev. Lett. {\bf 68} 1046; 
        1993 Phys. Rev.  {\bf B 48} 7297; 
        1991 Nucl. Phys. {\bf B360} 641;  
        1994 J. Phys. {\bf A27}  5375
\bibitem{Furusaki} A. Furusaki and N. Nagaosa, 1994 
         Phys. Rev. Lett. {\bf 72} 892
    
\bibitem{Wongaff}E. Wong and I. Affleck,  1994 Nucl. Phys. {\bf B417} 403
\bibitem{Satsv}P. de Sa and A. Tsvelik, 1995 Phys. Rev.  {\bf B52} 3067 

\bibitem{Ess}F.H.L. Essler, 1996 J.  Phys. {\bf A29}  6183

\bibitem{Skl}E.K. Sklyanin, 1988 J. Phys. {\bf A21}  2375
\bibitem{Mecnep}L.Mezincescu and R.I. Nepomechie, 
         1991 J. Phys. {\bf A24} L17; 
         1992 J. Phys. {\bf A25} 2533
\bibitem{Forkar}A. F\"orster and M. Karowski, 
         1993 Nucl.  Phys. {\bf B396} 611; 
         1993 Nucl.  Phys. {\bf B408} 512
\bibitem{Veggo}H.J.de Vega and A. Gonzalez-Ruiz, 
         1994 Nucl. Phys. {\bf B417} 553
\bibitem{Yuefh}R.H. Yue, H. Feng and B.Y. Hou, 
        1996 Nucl. Phys. {\bf B462}  167
\bibitem{Schulz} H. Schulz, 1985 J. Phys. {\bf C18} 581
\bibitem{Asakawa} H. Asakawa and M. Suzuki, 1995 J. Phys. {\bf A29} 225 

\bibitem{Degyue}T. Deguchi and R.H. Yue, 
        1996 {\sl preprint} OCHA-PP-84, 1997  cond-mat/9704138

\bibitem{Shiwad}M. Shiroishi and M. Wadati,
         1997  J. Phys. Soc. Japan {\bf 66 } 1

\bibitem{wilson}K. G. Wilson, 1975 Rev. Mod. Phys. {\bf 47} 773

\bibitem{nozieres} P. Nozi{\'e}res, J. Low Temp. Phys. {\bf 17} (1974) 31. 


\bibitem{Woypen}F. Woynarovich and K. Penc, 
         1991 Z. Phys. {\bf B85}  268

\bibitem{Tak} M. Takahashi, 
        1969 Prog. Theor. Phys. {\bf 42}  1098; 
        1972 Prog. Theor. Phys. {\bf 47}  69; 
        1974 Prog. Theor. Phys. {\bf 52} 103 
\bibitem{Skosal} S. Skorik and H. Saleur, 1995 J. Phys.  {\bf A28} 6605
\bibitem{Kapsko}A. Kapustin and S. Skorik, 1996 J. Phys. {\bf A 29} 1629 
\bibitem{Tsuyam}O. Tsuchiya and T. Yamamoto, 
        1997 {\sl preprint} UT-Komaba-97-4
\bibitem{Bedfra}G. Bed\"urftig and H. Frahm, 
        1997 {\sl preprint} cond-mat/9702227
\bibitem{Essfra}F.H.L. Essler and H. Frahm, 
        1997 {\sl preprint} cond-mat/9702234
\end{thebibliography}
\end{document}